\documentclass{an}
\usepackage{graphicx}
\usepackage{times}
\Volume{00}              
\Year{0000}              
\Month{00}               
\Pagespan{000}{000}      

\begin{document}

\newcommand{\kmsec}{\,\mbox{$\mbox{km}\,\mbox{s}^{-1}$}}
\newcommand{\phiorb}{$\phi_{\tiny orb}$}
\newcommand{\msun}{\hbox{$\hbox{M}_\odot$}}
\newcommand{\rsun}{\hbox{$\hbox{R}_\odot$}}
\newcommand{\ha}{\hbox{$\hbox{H}\alpha$}}
\newcommand{\hb}{\hbox{$\hbox{H}\beta$}}
\newcommand{\pb}{\hbox{$\hbox{P}\beta$}}
\newcommand{\hg}{\hbox{$\hbox{H}\gamma$}}
\newcommand{\hd}{\hbox{$\hbox{H}\delta$}}
\newcommand{\heii}{\hbox{$\hbox{He\,{\sc ii}\,$\lambda$4686\,\AA}$}}
\newcommand{\heia}{\hbox{$\hbox{He\,{\sc i}\,$\lambda$4472\,\AA}$}}
\newcommand{\heib}{\hbox{$\hbox{He\,{\sc i}\,$\lambda$6678\,\AA}$}}
\newcommand{\heic}{\hbox{$\hbox{He\,{\sc i}\,$\lambda$5876\,\AA}$}}

\lhead[\thepage]{A.N. Author: Title}
\rhead[Astron. Nachr./AN~{\bf XXX} (200X) X]{\thepage}
\headnote{Astron. Nachr./AN {\bf 32X} (200X) X, XXX--XXX}

\title{Doppler tomography of Cataclysmic Variables}

\author{L. Morales-Rueda}
\institute{University of Southampton, UK}
\date{Received {date will be inserted by the editor}; 
accepted {date will be inserted by the editor}} 

\abstract{The study of cataclysmic variables (CVs), and in particular
of the evolution of their accretion discs throughout their different
brightness states, has benefited largely from the use of indirect
imaging techniques. I report on the latest results obtained from
Doppler tomography of CVs concentrating mainly on results published
since the 2000 Astrotomography meeting in Brussels. Emphasis is given
to the spiral structures found in the accretion discs of some CVs, to
the evolution of these structures throughout quiescence and outburst,
and to our search for them in more systems.  }
\correspondence{lmr@astro.soton.ac.uk}

\maketitle

\section{Introduction}

Cataclysmic Variables (CVs) are close interacting binaries that
consist of a white dwarf (WD) as the compact object and either a main
sequence star or a slightly evolved star as the donor component. The
companion or donor transfers material to the more massive WD. Due to
conservation of angular momentum, and if the magnetic field of the WD
is not too strong, the mass being transferred forms an accretion disc
around the WD. If the magnetic field of the WD is strong (a few MGs)
there are two possible scenarios that can take place: a) the material
latches to the magnetic field lines before a disc can be formed and
accretion onto the WD occurs along the field lines in which case we
have a CV called a polar (Schwope, this volume), or b) the disc does
form but it gets disrupted in its inner regions by the magnetic field
lines in which case we have a CV called an intermediate polar.

The study of CVs started more than a century ago and many monographs
have been dedicated to their study, in particular the book by Warner
(1995) gives a very complete description of these systems.  The
components of a CV cannot be resolved directly as their angular size
from Earth is only a few tens of micro-arcseconds, thus the only way
to image CVs is by using indirect imaging techniques. Several indirect
imaging techniques have been used for their study throughout the
years, e.g. Eclipse Mapping, Physical Parameter Eclipse Mapping,
Stokes Imaging, Roche Tomography, Doppler Tomography etc. Doppler
Tomography (Marsh \&\ Horne 1988, Marsh 2001, Steeghs this volume)
uses a series of spectra covering the orbit of the binary and produces
2-dimensional maps of its velocity field. It is then possible to
transform this velocity map into a space map by making the assumption
that the accretion flow moves in a Keplerian fashion. We know from
observations that this is not always the case (e.g. Marsh \&\ Horne
1990) so it is usually preferable to study the velocity maps
themselves.

\section{Recent results from Doppler tomography}

The presence of a disc or ring of gas around the compact object in a
CV is usually obvious just by looking at high time resolution trailed
spectra covering most of an orbit. Emission lines in CVs are generally
more complicated than that, showing many components that arise in
different regions of the binary and that move at different
velocities. Thus it is usually not so obvious to assign these
components to emission regions in the CV. It is in these cases when
Doppler tomography plays an important role.

Thanks to Doppler tomography we have been able to identify the
presence of spiral structure in the accretion discs of CVs
(e.g. IP~Peg in outburst: Steeghs, Harlaftis \&\ Horne 1997; Harlaftis
et al. 1999; Morales-Rueda, Marsh \&\ Billington 2000, U~Gem in
outburst: Groot 2001) some ten years after it had been predicted
(Sawada, Matsuda \&\ Hachisu 1986). We have also been able to map the
accretion stream in magnetic CVs (Schwope, Mantel \&\ Horne 1997, and
other examples in this volume) and to map the irradiation of the donor
star (Morales-Rueda et al. 2000, Harlaftis 1999, Unda-Sanzana et
al. in press). Doppler tomography does not always solve the problem
though, in some cases we still do not know what the origin of some
components is, i.e. more cases of low velocity emission near the
centre of mass of the binary are appearing in the literature and in
most cases we do not have an explanation for it (North et al. 2001,
Unda-Sanzana, Marsh \&\ Morales-Rueda, in press).

\begin{figure}
\begin{center}
{\includegraphics[scale=0.48,angle=-90]{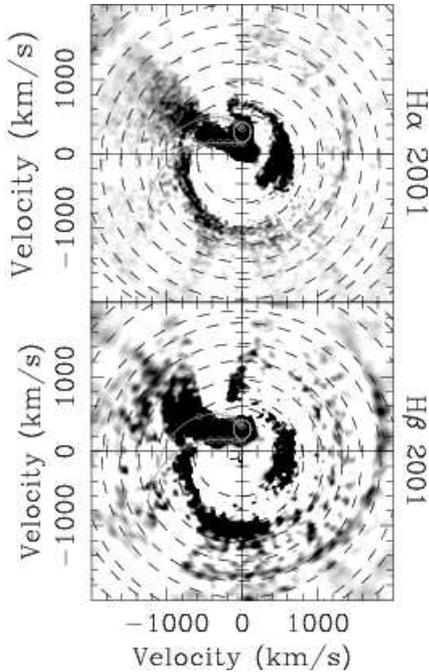}}
\end{center}
\caption{\ha\ and \hb\ Doppler maps of U~Gem during quiescence. After
  subtracting the elliptical background Unda-Sanzana et al. (in press),
  find signatures of spiral structure in the accretion disc.}
\label{fig1}
\end{figure}

Marsh (2001) gave a summary of all the Doppler maps published in the
literature since the start of tomography in astronomy. Here we extend
his summary by presenting in Table~\ref{table1} all the CVs that have
been mapped since then. A highlight of this summary are the tomograms
obtained during the rare (once every 33 years approximately)
superoutburst of WZ~Sge in 2001. These showed the presence of a strong
spiral structure during the start of the outburst, weakening as the
flux decays, appearing again during a re-brightening and finally
disappearing as the disc cools down. The maps also show the presence
of the donor star in the system which is visible due to an increase in
irradiation during the outburst. The detection of the donor star in
the Doppler maps allowed the determination of basic system parameters
such as the radial velocity amplitudes of both components. These
measurements helped constrain the masses of the donor star and the
white dwarf (Steeghs et al. 2001). More details on the 2001 outburst
of WZ~Sge are discussed by Steeghs (this volume) and references
therein. Other highlights include two maps of U~Gem one obtained at
outburst maximum and the other four days later showing that both
spiral arms evolve in a different way indicating asymmetric evolution
of the disc. The spiral structure not only varies in strength but also
in size and position in the disc (Boffin \& Steeghs 2002). Also of
interest are the maps of helium CVs like AM~CVn, that allowed the
authors to constrain the system parameters (Nelemans et al. 2001) and
GP~Com, that showed a rather complicated structure in the bright spot
emission (Morales-Rueda et al. 2003), maps of IP~Peg during quiescence
possibly showing the presence of spiral structure (Neustroev et
al. 2002), and maps of magnetic systems that allowed the authors to
constrain the accretion geometry (Schwope 2001).

\begin{table*}[h!]
\caption{List of Cataclysmic Variables for which Doppler tomograms
  have been obtained. The codes used in the table are the same as
  those in Marsh (2001). Types include: dwarf novae (DN), old nova
  (N), nova-like (NL), helium CV (HeCV), intermediate polar (IP),
  polar (P). States include: quiescence (Q) or outburst (O),
  superoutburst (SO), standstill (SS), flaring (F), high (H), low (L)
  and middle (M). The features found in the maps are rings (1), spots
  (2), the donor star (3), spiral structure (4), the gas stream (5)
  and low velocity emission (6). The four quadrants of the map are
  marked from ``a'' to ``d'' starting with the upper left quadrant and
  moving anticlockwise. Note 1: maps obtained on several nights during
  outburst are presented showing the evolution of the emission
  sites. Note 2: data obtained in the decline from outburst. Note 3:
  the gas stream emission does not follow a ballistic trajectory.}
\label{table1}
\begin{tabular}{lllllll}
\hline
Object & Type & State & Resolution & Line(s) & Features & References\\
       &      &       & km\,s$^{-1}$ &       &          &           \\
\hline
IYUMA & DN & Q & 220 & \ha, \hb, \heic & 1, 2a & Rolfe et al. (2001)\\
$''$    & $''$ & O & ?   & \ha & 1, 2abc, 3$^1$ & Rolfe et al. (2002)\\ 
WX Cet & DN & Q & 206 & \ha & 2cd, 3 & Tappert et al. (2003)\\
AK Cnc & DN & O & 91 & \ha &  6 & Tappert et al. (2003)\\
AQ Eri & DN & Q & 91 & \ha & 3, 6 & Tappert et al. (2003)\\
VW Hyi & DN & O/Q$^2$ & 50 & \ha & 1, 3, 5 & Tappert et al. (2003)\\
RZ Leo & DN & Q & 274, 206 & \ha & 2abcd & Tappert et al. (2003)\\
TU Men & DN & Q & 91 & \ha & 1, 3 & Tappert et al. (2003)\\
HS Vir & DN & Q & 114 & \ha & 2a, 6 & Tappert et al. (2003)\\
IP Peg & DN & Q & 125, 73  & \ha, \hb, \hg, \hd & 1, 2a, 2c, 4? & Neustroev et al. (2002)\\
FS Aur & DN & Q & 171 & \hb, \hg, \heia & 6 & Neustroev (2002)\\ 
$''$   &$''$&$''$&   $''$& \heii & 1 in He\,{\sc ii}? & Neustroev (2002)\\
WZ Sge & DN & O & 36 & \ha & 1, 3, 4? & Steeghs et al. (2001)\\
$''$     & $''$ & O & 100 & \heii & 4? & Baba et al. (2001)\\
$''$     & $''$ & O & 53 & \heii & 4 & Kuulkers et al. (2002)\\
$''$     & $''$ & O & ? & P$\beta$, P$\gamma$ & 2acd & Howell et al. (2003)\\
$''$     & $''$ & O & ? & He\,{\sc i}\,1.083$\mu$, He\,{\sc ii}\,1.163$\mu$ &
1, 2ab, 2cd & Howell et al. (2003)\\ 
BV Pup & DN & Q & 67 & \hb & \hb: 1, 2a & Bianchini et al. (2001)\\ 
$''$ &$''$  & $''$& $''$ & \heii & He\,{\sc ii} 2b & Bianchini et al. (2001)\\ 
EM Cyg & DN & SS & 36 & \ha & 1, 3 & North et al. (2001, 2002)\\
V426 Oph & DN & Q & 36 & \ha & 6 & North et al. (2001, 2002)\\
SS Cyg & DN & Q & 36 & \ha & 3, 6 & North et al. (2001, 2002)\\
AH Her & DN & Q & 36 & \ha & 3, 6 & North et al. (2001, 2002)\\
U Gem & DN & O & 30 & \heii & 1, 3, 4 & Boffin \& Steeghs (2002)\\
$''$ & $''$ & Q & 13 & \heii & 2a & Unda-Sanzana \& Marsh (2002)\\
$''$ & $''$ & Q & 18 & \heib & 2a, 3, 6 & Unda-Sanzana \& Marsh (2002)\\
OY Car & DN & O & ? & \ha & 3, 5 & Mason \& Howell (2002)\\
EX Hya & IP & - & ? & \ha & 1, 2ab & Wynn (2001)\\
V1025 Cen & IP & - & 86 & \hb & 1, 3, 5 & Hellier et al. (2002)\\
UZ For & P & - & ? & \heii & 3, 5 & Schwope (2001)\\
V1309 Ori & P & - & 115 & \hg, \heia & 3, 5 & Staude et al. (2001)\\
 $''$ & $''$& $''$& 115 & \heii & 3, 5 & Staude et al. (2001)\\
 $''$ & $''$& $''$& 84 & He\,{\sc i}\,$\lambda$8236\,\AA & 3, 5 & Staude et al. (2001)\\
BY Cam & P & - & ? & \hb & 6 & Schwope (2001)\\
V1432 Aql & P & - & ? & \heii & 6 & Schwope (2001)\\
RXJ1313 & P & H & 91 - 137 & \ha & 3 & van der Heyden et al. (2002)\\  
$''$&$''$& $''$& 133 & \hb, \hg, \heii & 3, 5 & van der Heyden et al. (2002)\\ 
V834 Cen & P & - & ? & \heii & 3, 5 & Potter et al. (2001)\\
AM Her & P & - & 80 & \heii & 3, 5$^3$ & Schwarz et al. (2002)\\
$''$&$''$& $''$& 50 & Na\,{\sc i}, Ca\,{\sc ii} & 3 &  Schwarz et al. (2002)\\
UW Pic & P & - & 93 & \hb, \heii & 3, 5 & Romero-Colmenero et al. (2003)\\
V895 Cen & P & H & 229 & \ha & 3, 5? & Salvi et al. (2002)\\
V841 Oph & N & - & 82 - 91& \ha, \heib & 1?, 3, 6 & Diaz \&\ Ribeiro (2003)\\
RR Pic & N & - & 112 & \ha, \heib & 1, 2ac & Schmidtobreick et al. (2003) \\
GP~Com & HeCV & - & 56(red), 38(blue) & HeI, HeII & 1, 2a, 6 & Morales-Rueda et al. (2001, 2003)\\
AM CVn & HeCV & Q & 138 &  \heia & 1, 2a & Nelemans et al. (2001)\\
\hline
\end{tabular}
\end{table*}

\subsection{Doppler tomography of discs: spiral structure}

Since their observational discovery in 1996 astronomers have found
spiral structure in the accretion discs of six CVs. IP~Peg, U~Gem and
WZ~Sge are the best studied systems of the six with spiral structure
having been detected in several different sets of data. In the case of
WZ~Sge the spectra were all taken during the same high state (as this
system only outbursts every 20-30 years) whereas for IP~Peg the
structure has been seen in three different outbursts and for U~Gem in
two. For IP~Peg and U~Gem some authors have also found indication of
spiral structure during quiescence (IP~Peg: Neustroev et al. 2002,
U~Gem: Unda-Sanzana et al., in press, see Fig.~\ref{fig1}). The other
three CVs that show spiral structure are SS~Cyg, EX~Dra and
V347~Pup. All, apart from V347~Pup which is a nova-like, are dwarf
novae.

The presence of spiral structure in the accretion disc is of great
importance as spiral shocks have been called upon as a possible way to
get rid of angular momentum in the accretion disc. Magnetorotational
instabilities (Hawley \&\ Balbus 1995) are the widely accepted way to
explain the effective viscosity in outburst discs but there still is
some debate regarding angular momentum transport in quiescent discs.
Hydrodynamic tidal stresses (Sawada, Matsuda \&\ Hachisu 1986) that
result in spiral shocks could also contribute to the viscosity
together with magnetorotational instabilities.

More recently, Smak (2001) and Ogilvie (2002) have interpreted these
spiral structures in the accretion disc as vertical enhancements of
the disc being irradiated by the white dwarf. The predictions of this
tidal model are viscosity-dependent, therefore Smak's and Ogilvie's
model has the potential to measure the viscosity in discs.

Shocks and vertical enhancements would evolve in different ways when
the disc cools down and heats up during quiescence and outburst
respectively. In the case of shocks, as the Mach numbers in the disc
increase during quiescence, we would expect the opening angle of the
spirals to decrease, producing very wound up spirals as opposed to the
large opening angle spirals seen during outburst. According to Ogilvie
(2002), in the case of vertical structure caused by tidal interactions
with the donor star, the cooling down of the disc only contributes to
decreasing the luminosity of the central regions of the CV and the
irradiation of the tidal distortions. The spiral structure will appear
fainter but it would not move its position in the disc. During
quiescence accretion discs are also known to shrink (e.g. Wood et al.
1989) thus if the radius of the disc becomes smaller than the tidal
radius, the vertical structure will not form.

At present there seem to be arguments in favour and against both
explanations. For example, in favour of the shocks we have Baptista's
(this volume) signatures of sub-Keplerian velocity in the disc in
eclipse maps of IP~Peg during outburst. We would expect the gas to
slow down as it reaches the shocks, which is what these sub-Keplerian
velocities seem to be indicating. This is not to be confused with the
sub-Keplerian velocities seen in the outer discs of some CVs which can
be either produced by the collision of the accretion stream with the
disc, or intrinsic to the outer disc, e.g. Marsh 1988. On the other
hand, Neustroev et al. (2002) and Unda-Sanzana et al. (in press) find
indications of the presence of spiral structure in the discs of IP~Peg
and U~Gem respectively {\it during quiescence} (see
Fig.~\ref{fig1}). If confirmed, this would indicate that the spiral
structure is not a wave or a shock that evolves to smaller opening
angles when the disc cools down, but the result of tidally thickened
sectors of the disc being irradiated by the white dwarf, boundary
layer and/or inner disc. During quiescence the luminosity of the
central regions of the disc will be smaller and the spiral pattern
will be fainter than during outburst.

Currently there are two data sets known to the author that cover a
large part of the outburst of two CVs known to show spiral structure
in their discs. The first one consists of 5 nights taken at the INT
and several at the MMT during the April/May 2001 outburst of U~Gem
(Steeghs et al. in preparation). The second one corresponds to the
2001 outburst of WZ~Sge and is discussed by Steeghs (this
volume). From these two data sets we can see how the spiral structure
changes from night to night during the outburst. In the case of U~Gem
the last two nights of data were taken when the system was declining
from outburst. The spirals are no longer present in the spectra taken
in these two nights. Unfortunately, due to large gaps in the coverage
of the quiescence-outburst-quiescence transitions none of these data
sets allow us to see whether the spirals wind themselves up or remain
the same but just go fainter.

We envisage that the only way to study in detail these state
transitions and therefore to discern between the two models proposed
to explain the spirals, is by making use of the target of opportunity
and monitoring capabilities of a robotic telescope. We believe that
the 2\,m Liverpool telescope + spectrograph (Morales-Rueda et al.,
this volume) will be the best combination to obtain the data that will
allow us to answer these questions.

\section{Conclusions}

Doppler tomography has proved to be a very powerful technique to study
CVs. Recent results include the detection of the donor star in WZ~Sge
during outburst, the detection of spiral structure in WZ~Sge during
outburst and in IP~Peg and U~Gem during quiescence.

We believe that Doppler tomography will be the key to understanding
the evolution of spiral structure in accretion discs during the
accretion state transitions and therefore will help us distinguish
between the models proposed for the formation of this structure with
important consequences regarding the viscosity in outburst discs.

\acknowledgements

The author would like to thank E. Unda-Sanzana for providing Fig.~1.

\end{document}